\documentclass[12pt,indent]{article}
\usepackage{a4,latexsym,amsmath,amssymb}
\usepackage[latin1]{inputenc}
\usepackage{float}
\usepackage{natbib}
\usepackage{graphics}
\usepackage[english]{babel}
\usepackage{tabularx}
\usepackage{lscape}
\usepackage{multirow}
\usepackage{rotating}
\usepackage{dsfont}
\usepackage{caption}
\usepackage{subcaption}
\usepackage{bbm}
\usepackage[table]{xcolor}
\usepackage{booktabs}
\usepackage{calc}
\usepackage{ifthen}
\usepackage{url}
\usepackage{colortbl}      
\usepackage{tikz}

\usepackage{blindtext}
\usepackage{scrextend}
\usepackage{mathtools}
\usepackage{verbatim}
\usepackage{setspace}
\usetikzlibrary{shapes,snakes,matrix}

\newenvironment{tabularsmall}
{ \footnotesize \sffamily \tabular } {
\endtabular
\normalfont }

\newcommand{\E}{\operatorname{E}}      

\newcommand{\deltab}{\boldsymbol{\delta}}

\newcommand{\thetab}{\boldsymbol{\theta}}

\newcommand{\blanco}[1]{}

\def\d{\displaystyle}

\usepackage{natbib}

\usepackage{geometry}
\usepackage{setspace}
\usepackage{tikz}
\usepackage[]{graphicx}
\usepackage[]{color}
\makeatletter
\def\maxwidth{ %
  \ifdim\Gin@nat@width>\linewidth
    \linewidth
  \else
    \Gin@nat@width
  \fi
}
\makeatother

\definecolor{fgcolor}{rgb}{0.345, 0.345, 0.345}

\usepackage{framed}
\makeatletter
 {\par\unskip\endMakeFramed%
 \at@end@of@kframe}
\makeatother

\definecolor{shadecolor}{rgb}{.97, .97, .97}
\definecolor{messagecolor}{rgb}{0, 0, 0}
\definecolor{warningcolor}{rgb}{1, 0, 1}
\definecolor{errorcolor}{rgb}{1, 0, 0}

\usepackage{alltt}
\IfFileExists{upquote.sty}{\usepackage{upquote}}{}

\usepackage{amsthm}
 
\theoremstyle{definition}
\newtheorem{definition}{Definition}[section]

\newtheorem{theorem}{Proposition}[section]

\begin{document}
\bibliographystyle{chicago}
\sloppy

\makeatletter
\renewcommand{\section}{\@startsection{section}{1}{\z@}%
        {-3.5ex \@plus -1ex \@minus -.2ex}%
        {1.5ex \@plus.2ex}%
        {\reset@font\large\sffamily}}
\renewcommand{\subsection}{\@startsection{subsection}{1}{\z@}%
        {-3.25ex \@plus -1ex \@minus -.2ex}%
        {1.1ex \@plus.2ex}%
        {\reset@font\normalsize\sffamily\flushleft}}
\renewcommand{\subsubsection}{\@startsection{subsubsection}{1}{\z@}%
        {-3.25ex \@plus -1ex \@minus -.2ex}%
        {1.1ex \@plus.2ex}%
        {\reset@font\normalsize\sffamily\flushleft}}
\makeatother



\newsavebox{\tempbox}
\newlength{\linelength}
\setlength{\linelength}{\linewidth-10mm} \makeatletter
\renewcommand{\@makecaption}[2]
{
  \renewcommand{\baselinestretch}{1.1} \normalsize\small
  \vspace{5mm}
  \sbox{\tempbox}{#1: #2}
  \ifthenelse{\lengthtest{\wd\tempbox>\linelength}}
  {\noindent\hspace*{4mm}\parbox{\linewidth-10mm}{\sc#1: \sl#2\par}}
  {\begin{center}\sc#1: \sl#2\par\end{center}}
}



\def\R{\mathchoice{ \hbox{${\rm I}\!{\rm R}$} }
                   { \hbox{${\rm I}\!{\rm R}$} }
                   { \hbox{$ \scriptstyle  {\rm I}\!{\rm R}$} }
                   { \hbox{$ \scriptscriptstyle  {\rm I}\!{\rm R}$} }  }

\def\N{\mathchoice{ \hbox{${\rm I}\!{\rm N}$} }
                   { \hbox{${\rm I}\!{\rm N}$} }
                   { \hbox{$ \scriptstyle  {\rm I}\!{\rm N}$} }
                   { \hbox{$ \scriptscriptstyle  {\rm I}\!{\rm N}$} }  }

\def\d{\displaystyle}\def\d{\displaystyle}

\title{What is an Ordinal Latent Trait Model?}
  \author{Gerhard Tutz \\{\small Ludwig-Maximilians-Universit\"{a}t M\"{u}nchen}\\{\small Akademiestra{\ss}e 1, 80799 M\"{u}nchen}}


\maketitle

\begin{abstract} 
\noindent
Although various polytomous item response models are considered to be ordinal models there seems no general definition of an ordinal model available.
Alternative concepts of ordinal models are discussed and it is shown that they coincide for classical unidimensional models.  For multidimensional models the definition of an ordinal model refers to specific traits in the multidimensional space of traits. The objective is to provide a theoretical framework for ordinal models. Practical considerations concerning the strength of the link between the latent trait and the order of categories are considered  briefly.

\end{abstract}

\section{Introduction}

Various latent trait models for ordered response data are in common use,  for overviews see, for example, \citet{ostini2006polytomous} and \citet{VanderLind2016}.  In particular   three basic types of models, cumulative models, sequential models and adjacent categories models have been proposed. 
All these models are considered as using the ordinal scale information but surprisingly there seems no common framework available that tells us when a model should be considered as an ordinal model and when not.  

For the characterization one might think of invariance properties. They
may certainly be used  to characterize \textit{nominal} item response models. One can consider an item response model to be nominal if it is stable under 
any permutation $T:\{0,1,\dots,k\}\longmapsto\{0,1,\dots,k\}$, where $\{0,1,\dots,k\}$ are the response categories.  That means, the same model holds  if one permutes the categories, see also \citet{peyhardi2014new}. However, the definition   distinguishes between nominal and non-nominal models only. A non-nominal model is not automatically an ordinal model. It does not necessarily exploit the ordering of categories. One might also think of characterizing ordinal models by using the  reverse  permutation $p(r)=k-r$, $r=0,\dots,k$. Of course a model that is stable under the reverse  permutation but not under other permutations uses in some sense the order in the categories, but why should an ordinal model be stable under the reverse permutation? Several models like the graded response model to be considered later are indeed stable under the reverse permutation if the link function is symmetric. However, they also exploit the order  of categories if the link function is not symmetric, though not being   stable under the reverse permutation.

A  specific branch of the literature investigates measurement properties of polytomous item response models that are linked to the ordering of categories.
For example,  \citet{hemker2001measurement} consider  measurement properties of continuation ratio models and examine concepts like invariant item ordering, see also \citet{sijtsma1998nonparametric}, \citet{van2005stochastic}. They do not aim at defining what an ordinal model is but reflect order properties. Although the findings are generally helpful, in some cases  questionable tools are used, for example, when   the expectation of the categorical response is used to define order it is implied that the response is measured on an interval scale level.

In the following we consider alternative definitions of an ordinal model. All of them link the ordering of categories or groups of categories to the order of the   latent trait.   Although the focus is on unidimensional models,  multidimensional models are considered briefly. For them the characterization as ordinal may refer to specific traits. 
 
\section{Ordinal Item Response Models}

Let the response of person $p$ on one specific item be given by $Y_{p} \in \{0,1,\dots,k\}$. 
An item response model is in general determined by a parameterization of the response probabilities 
\begin{equation}\label{eq:gen0}
\pi_{r}(\theta_p)= {P(Y_{p} = r|\theta_p, \deltab )},\quad r=0,\dots,k,
\end{equation}
where $\theta_p$ is a person or trait parameter and $\deltab$ is a vector of item parameters. For simplicity we will mostly suppress  the dependency on the item parameter  in the notation.

A specific ordinal model, which is in the following used for illustrative purposes, is   Samejima's \textit{graded response model},  
\begin{equation}\label{eq:cum1}
P(Y_{p} \ge r |\theta_p) = F(\theta_p-\delta_{r}),\quad r=1,\dots,k.
\end{equation}
where $F(.)$ is a strictly monotonically increasing distribution function and $\deltab^T=(\delta_{1},\dots,\delta_{k})$. If one uses for $F(.)$ the logistic distribution one obtains the Rasch type version of the graded response model. The model is from the class of \textit{cumulative models} because  on the left hand side one has the cumulative probabilities of categories $\{r,\dots,k\}$. It  can also be seen as a collection of binary models. If one considers the grouped response $Y_{p}^{(r)}=1$ if $Y_{p} \in \{r,\dots,k\}$ and $Y_{p}^{(r)}=0$, otherwise, the model specifies that $Y_{p}^{(r)}$ follows a binary response model. Of course the binary models have to hold simultaneously and contain the same person parameter $\theta_p$, which makes the model uni dimensional in terms of person parameters. If $F(.)$ is the logistic distribution the partitioning into categories $Y_{p} < r$ and $Y_{p} \ge r$ is determined by the binary Rasch model with item parameter $\delta_{ir}$. 

In all cumulative models item parameters have to be ordered, that is, $\delta_{1}\le\dots\le\delta_{k}$. If two item parameters coincide response probabilities of specific categories become zero, if, for example, $\delta_{1}=\delta_{2}$ one has $P(Y_{p} = 1 |\theta_p)=0$. In the following we exclude the degenerate case, that is, we assume that all probabilities are positive, however small. 

\subsection{General Criteria for Ordinal Models}

The basic assumption made here is that the response categories $\{0,1,\dots,k\}$ are ordered. A model can be considered  an ordinal item response model if it explicitly uses  the order of categories. An attempt to define the order  of categories was given by \citet{adams2012rasch}. They consider categories in an item response model as ordered if the expectation $\E(Y_{p}|\theta_p)$ is an increasing function of $\theta_p$. That means, if a person has a higher value of $\theta_p$ than another person, then the person with the higher value will, on average, score more.  The problem with the definition is that the expectation is a sensible measure only if the response $Y_{p}$ is measured on a metric scale level. Moreover, the starting point here is that $Y_{p}$ is measured on an ordinal scale level. In contrast to \citet{adams2012rasch}  we do not want to \textit{define ordering of categories} but investigate if a model \textit{uses} the order of the response categories. The expectation is   not helpful for this purpose because it uses a scale level that is not available.

Instead of using the expectation it seems more appropriate  to consider the response probabilities themselves. 
Somewhat informal an item response model can be considered an ordinal model if   an increase in the trait parameter has the effect that the probabilities of higher response categories grow and the probabilities of lower response categories decrease. However, to obtain a concise conceptualization one has to specify what growing and decreasing of probabilities is supposed to mean.  A criterion that is inspired by the cumulative model is the following. 
\vspace{-0.5 cm}
\begin{addmargin}[0.4 cm]{-0.0 cm}
 \begin{definition}\label{df:cum}
An item response model is considered an ordinal model according to the \textit{split concept} if one has  for $r=1,\dots,k$ 
\begin{equation}\label{eq:cum}
 P(Y_{p} \ge r |  \theta_p) \text{ is an increasing function  of } \theta_p. 
\end{equation}
\end{definition}
\end{addmargin}
The definition clearly uses the ordering of categories. It partitions the response categories into the sets $\{0,\dots,r-1\}$ and $\{r,\dots,k\}$, which makes  sense if categories are ordered only. In the definition the increase in probabilities for increasing $\theta_p$ refers to the whole set of higher categories $\{r,\dots,k\}$.  We refer to Definition  \ref{df:cum} as the split concept of an ordinal model since it is based on the splitting of the set of categories.
It is obvious that cumulative models of the form (\ref{eq:cum1}) are ordinal in the sense of this definition.

For clarity it should be noted that we call a function  increasing if it is strictly monotonically increasing. Of course one could also use a weaker concept that postulates that the function is merely non-decreasing. The stronger concept based on increasing functions has the advantage that it postulates that the trait is definitely linked to the response probabilities.    

Property (\ref{eq:cum}) can also be seen as a \textit{stochastic ordering} property since it implies that for $\theta_{p_1}< \theta_{p_1}$ one has 
\[
P(Y_{p} \ge r |  \theta_{p_1}) < P(Y_{p} \ge r |  \theta_{p_2}) \text{ for all } r.
\]
It is a stochastic ordering property that refers to one item, in contrast to ordering properties considered, for example, by \citet{hemker2001measurement}, where the ordering refers to sums of scores  over items.

Splitting the categories into two sets of categories is  an option if one wants  to account for order, however it is certainly not the only one. The order of categories can also be reflected by using a pair of categories instead of sets of categories. This order  requirement can be formalized as follows.

\vspace{-0.5 cm}
\begin{addmargin}[0.4 cm]{-0.0 cm}
\begin{definition}\label{th:adj}
An item response model is considered an ordinal model according to the \textit{paired categories concept} if   for any pairs of categories $(s,r), s <r$ 
\begin{equation}\label{eq:Ac}
 \frac{P(Y_{p} = r |  \theta_p)}{P(Y_{p} = s |  \theta_p)} \text{ is an increasing function  of } \theta_p. 
\end{equation}
\end{definition}
\end{addmargin}
 According to this definition one expects in an ordinal model that for any pair of  categories the probability of the higher category increases as a function of the trait.  The definition explicitly uses pairs of categories, therefore the name paired categories concept. 

It is not hard to show that one can alternatively postulate 
\begin{align}\label{eq:Ac2}
 P(Y_{p} = r | Y_{pi} \in \{s,r\},\theta_p ) &\text{ is an increasing function  of } \theta_p   \\\nonumber
 &\text{ for any pairs of categories } (s,r), s <r.
\end{align}
or restrict the postulate to adjacent categories 
\begin{equation}\label{eq:Ac3}
 P(Y_{p} = r | Y_{pi} \in \{r-1,r\},\theta_p ) \text{ is an increasing function  of } \theta_p \text{ for  } r=1,\dots,k.
\end{equation}
All these definitions are equivalent and yield the paired categories concept of an ordinal model. 
The representation (\ref{eq:Ac2}) was also used by \citet{adams2012rasch} as a  definition of order of categories.


While the  split concept is based on groups of variables, the paired categories concept uses single categories. A third type of monotonicity compares single 
categories to groups of categories conditionally.
\vspace{-0.5 cm}
\begin{addmargin}[0.4 cm]{-0.0 cm}
\begin{definition}\label{th:Sequ}
An item response model is considered an ordinal model according to the \textit{conditional concept} if    for $r=1,\dots,k$ 
\begin{equation}\label{eq:sequ}
 P(Y_{p} \ge r | Y_{p} \ge r-1, \theta_p) \text{ is an increasing function  of } \theta_p. 
\end{equation}
\end{definition}
\end{addmargin}
 That means given the response is in categories \{r-1,\dots,k\}  the probability of the  higher categories \{r,\dots,k\}  increases as a function of the trait. Thus one compares if the response is in category $r-1$ or in the categories \{r,\dots,k\} given the response is at least in category $r-1$.

 All three concepts  link the concept of an ordinal model to the monotonicity of probabilities of specific categories or groups of categories that are built by taking  into account the order of categories. Although one might suspect that they are equivalent, in general they are not. However, there is some hierarchy in the concepts. It turns out that the conditional concept is the strongest followed by the paired categories concept.  The split concept is the weakest one.  
 As shown in the appendix the following proposition holds.

\vspace{-0.5 cm}
\begin{addmargin}[0.4 cm]{-0.0 cm}
\begin{theorem}\label{th:gen}
For a multi-category item response model one obtains:
\begin{itemize}
\item[(1)] If a model is ordinal according to the paired categories concept it is also  ordinal according to the conditional concept (if (\ref{eq:Ac2}) holds also (\ref{eq:sequ}) holds).

\item[(2)] If a model is ordinal according to the conditional concept it is also  ordinal according to the split concept  
(if (\ref{eq:sequ}) holds also (\ref{eq:cum}) holds).
\end{itemize}
\end{theorem}
\end{addmargin} 
In short form one has
\[
(\text{paired} ) \Rightarrow (\text{conditional} ) \Rightarrow (\text{split} ).
\]

The concepts are not equivalent. One can  construct  item response models that are ordinal according to the conditional concept but not according to the paired categories concept. Also item response models can be ordinal according to the split concept but not to the conditional concept.

\subsection{Monotonicity in Classical Item Response Models}

In the previous section several conceptualizations of a ordinal response model have been given. In the following it is investigated if classical response models that are typically considered ordinal models meet these requirements. First the models are briefly sketched. 

\subsubsection*{Adjacent Categories Models }
The general \textit{adjacent categories} model has the form 
\begin{equation}\label{eq:part}
P(Y_{p} = r | Y_{pi} \in \{r-1,r\}, \theta_p ) = F(\theta_p-\delta_{ir}),\quad r=1,\dots,k,
\end{equation}
where $F(.)$ is a strictly monotonically increasing distribution function. \textit{Given} the response is in the adjacent categories   $\{r-1,r\}$ the probability of observing the higher response category is specified by a binary response model.

The most frequently used adjacent categories model  is the \textit{polytomous Rasch model}, also known as \textit{partial credit model}. It results from using the logistic distribution function $F(.)$, and has the simpler representation
\begin{equation}\label{eq:part0}
\log(\frac {P(Y_{p} = r |\theta_p)}{  P( Y_{pi} =r-1|\theta_p)}) = \theta_p-\delta_{ir},\quad r=1,\dots,k.
\end{equation}
The partial credit model was proposed by  \citet{Masters:82}, \citet{MasWri:84}, extensions were  given by \citet{glas1989extensions}, \citet{von2004partially}.  
The polytomous Rasch model representation was investigated, among others, by \citet{andrich1978rating}, \citet{Andrichh2016}.

\subsubsection*{Sequential Models}
The general \textit{sequential model} has the form 
\begin{equation}\label{eq:seq}
P(Y_{pi}\ge r|Y_{pi}\ge r-1, \theta_p)=F(\theta_p-\delta_{ir}), \quad r=1,\dots,k.
\end{equation}
It can be derived as a process model when   an item has to be solved in several steps with the scores representing the levels of the solution. 
In each of the consecutive steps a  binary response model applies.  In the first step the respondent tries to master the transition from level 0 to level 1. The corresponding model is $P(Y_{pi}^{(1)}=1)=F(\theta_p-\delta_{i1})$.
If the person does not master the first step, the process stops and one observes $Y_{pi}=0$. Otherwise,   $Y_{pi} \ge 1$ and the person tries to solve the next sub problem or step. In general, the $r$-th step, which is only tried if the previous steps were successful,  is determined by (\ref{eq:seq}).
For more details, see \citet{molenaar1983item}, \citet{Tutz:90b}, \citet{verhelst1997steps}, \citet{hemker2001measurement}.

\subsubsection*{Ordinality of Models}

Let us begin with adjacent categories models. Since the response function $F(.)$ is a strictly monotonically function all adjacent categories are ordinal according to the paired comparison concept. Since this is the strongest concept they are also ordinal according to the weaker concepts. The sequential model is obviously ordinal according to the conditional concept and therefore to the split concept. Cumulative models are ordinal according to the split concept. Table 
\ref{tab:overconc} gives an overview. The structure in Table \ref{tab:overconc} is a consequence of the definitions of an ordinal model, which are built by using the same comparisons of categories that are used when constructing models.

\begin{table}[h!]
\centering
\caption{Traditional models as ordinal models.}\label{tab:overconc}
\begin{tabularsmall}{@{}lccc@{}}
\toprule \medskip
               & Paired Categories Concept  &     Split Concept      &  Conditional  Concept             \\
               \midrule\medskip
{  Adjacent Categories Model  }     &  Yes     &   Yes                & Yes                               \\
\medskip
{  Sequential Model  }  &         &       Yes          &Yes                                \\
\medskip
{  Cumulative Model  }  &        &                   &Yes                               \\

\bottomrule
\end{tabularsmall}
\end{table}

One obtains a stronger result for specific versions of classic item response models.
The most common versions  are the logistic versions, which in the case of the adjacent categories model yields the partial credit model.
As shown in the Appendix   the logistic versions of all three model types  meet the requirements of all three definitions.

\vspace{-0.5 cm}
\begin{addmargin}[0.4 cm]{-0.0 cm}
\begin{theorem}\label{df:cum2}
The partial credit model, the logistic cumulative model and the logistic sequential  model are ordinal according to the paired categories concept and therefore also to the conditional and split concept.
\end{theorem}
\end{addmargin} 

\section{Alternative Models}
\subsection{IR-Tree Models}
Recently IR-tree models, which assume a nested structure with the building blocks given as binary models, have been proposed, see, for example, 
\citet{de2012irtrees},  \citet{bockenholt2013modeling}, \citet{khorramdel2014measuring},   \citet{bockenholt2016measuring} and \citet{bockenholt2017response}.
As an example we use a simple IR-tree model   following   the presentation of IR-tree models given by \citet{bockenholt2016measuring}. 
IR-tree models are sequential process model, a response is constructed based on a series of mental questions that are modeled as binary response models.

Let four response categories be given by ``strongly disagree (0)'',  ``weakly disagree (1)'',  ``weakly agree (2)'', ``strongly agree (3)''. A simple IR-tree model assumes that the distinction between agreement and disagreement categories is determined by the binary model  
\begin{equation}\label{eq:tree}
P(Y_{pi} \in \{2,3\}|\theta_p)=F(\theta_p-\delta^{(1)}),
\end{equation}  
and the conditional responses are determined by the binary models  
\begin{align}\label{eq:tree2}
&P(Y_{pi} =1|Y_{pi} \in \{0,1\}, \theta_p^{(2)})=F(\theta_p^{(2)}-\delta^{(2)}),\\\nonumber &P(Y_{pi} =3|Y_{pi} \in \{2,3\}, \theta_p^{(3)})=F(\theta_p^{(3)}-\delta^{(3)}).
\end{align} 
The first query determines a respondent's agreement or disagreement and is determined by model (\ref{eq:tree}). The second queries modeled in (\ref{eq:tree2})  determine  the extremity of the (dis)agreement. It is noteworthy that   each query has its own person and item parameter and response probabilities of single categories can depend on all the person parameters $P(Y_{pi} =r|\theta_p,\theta_p^{(2)}, \theta_p^{(3)} )$. 
A consequence is that the model is not ordinal with respect to the latent trait $\theta_p$.
For example, the proportion  $P(Y_{pi} =1|\theta_p,\theta_p^{(2)}, \theta_p^{(3)} )/P(Y_{pi} =0|\theta_p, \theta_p^{(2)}, \theta_p^{(3)} )$ is not a function of $\theta_p$  and therefore can not increase in $\theta_p$.

Therefore,  IR-tree models are \textit{not} ordinal models in the sense considered here. Only parts of the models could be considered ordinal, for example, how the model distinguishes between agreement and disagreement categories.

\subsection{Polytomous Rasch Model}

The unidimensional Rasch model has been considered by various authors, among them \citet{Rasch:61}, \citet{bock1972estimating}, \citet{Andersen:77}, \citet{roskam1989conditions}, \citet{Andrichh2016}. In Bock's   version it has the form
\[
P(Y_{pi}=r)= \frac{\exp(\alpha_{ir}(\theta_p-\beta_{ir}))}{\sum_{s=0}^{k}\exp(\alpha_{is}(\theta_p-\beta_{is}))}, \quad r=1,\dots,k.
\]
When using the paired categories concept one obtains 
\[
\log(\frac{P(Y_{p} = r |  \theta_p)}{P(Y_{p} = s |  \theta_p)})= ({\alpha_{ir}-\alpha_{is}})\theta_p-(\alpha_{\alpha_{ir}-\alpha_{is}})(\beta_{ir}-\beta_{is}),
\]
which is monotonically increasing only if ${\alpha_{ir}-\alpha_{is}}> 0$. Therefore, it is ordinal only if the slope parameters are ordered, that is, ${\alpha_{i1}\dots<\alpha_{is}}$. Although Bock's model is usually referred to as a nominal categories model the postulated uni-dimensionality together with the ordering of the slope parameters makes it an ordinal model. A truly nominal version is the multidimensional extension proposed by \citet{thissen2010nominal}.

\section{Multidimensional IRT Models}

In the previous section the concept of an ordinal model has been considered for uni-dimensional models. Although   the focus is on uni-dimensional models we briefly consider   multidimensional IRT models since for them also the question arises when is a model ordinal and in what sense. Multidimensional IRT models have been a vivid area of research, see, for example, \citet{kelderman1996multidimensional}, \citet{wetzel2015multidimensional}, \citet{plieninger2016mountain},  for partial credit type extensions and \citet{bolt2009addressing},
\citet{bolt2011multiscale}, \citet{FalkCai2016} for nominal models.

A partial credit type multidimensional model has the form
\[
\log \left(\frac {P(Y_{p}=r)}{P(Y_{p}=r-1)}\right)= \sum_{d=1}^D\omega_{rd}\theta_{pd}-\delta_{r},\quad r=1,\dots,k,
\]
where the person parameter $\thetab_{p}^T=(\theta_{p1},\dots,\theta_{pD})$ is now a vector with $D$ dimensions, and $\omega_{ird}$ are scoring weights. If 
$D=1$,   $\omega_{ird}=1$, one obtains the simple partial credit model. \citet{kelderman1996multidimensional} proposed to use discrete non-negative values $1,2,\dots$
for the scoring weights $\omega_{ird}$. 

For the definition of an ordinal model it is crucial that the person vector $\thetab_{p}$ contains several dimensions. Therefore it seems sensible to   
refer to the component in the definition. 
\vspace{-0.5 cm}
\begin{addmargin}[0.4 cm]{-0.0 cm}
\begin{definition}\label{th:adj}
An item response model is considered an ordinal model for $\theta_{pj}$ according the the paired categories concept if   for any pairs of categories $(s,r), s <r$ 
\begin{equation}\label{eq:Ac}
 \frac{P(Y_{p} = r |  \thetab_{p})}{P(Y_{p} = s |  \thetab_{p})} \text{ is an increasing function  of }  \theta_{pj}. 
\end{equation}
\end{definition}
\end{addmargin}

The definition links the ordinality of a model to a specific dimension. One could also think of postulating that the proportion of probabilities should be an increasing function for all dimensions. However, that would be much too restrictive and exclude models that link only one or two of the dimensions  
to the probabilities of categories.

Let us consider as an example a version of the partial credit model that includes response styles (\citet{TuSchBe2018}). In the   partial credit model with response style (PCMRS) one has  a model with two dimensions, which (for an odd number of response categories) has the form
\begin{equation}
\log \left(\frac {P(Y_{p}=r)}{P(Y_{p}=r-1)}\right)= \theta_{p1}+ (m-r+0.5)\theta_{p2} - \delta_{r}, \quad r=1,\dots,k,
\end{equation}
where $m=k/2$ denotes the middle category. The first  parameter $\theta_{p1}$ represents the trait to be measured whereas the second parameter   is a response style parameter, which  represents the tendency to middle or extreme categories.

It is obvious that the model is ordinal in the first dimension. It is not ordinal in  the second dimension. Let us consider as an example $k=4$. Then one has
$\log ({P(Y_{p}=1 }/{P(Y_{p}=0))} = \theta_{p1}+ 1.5\theta_{p2} - \delta_{r}$ and $\log ({P(Y_{p}=4 }/{P(Y_{p}=3))} = \theta_{p1}- 1.5\theta_{p2} - \delta_{r}$. Therefore, when comparing categories 1 and 0 one has an increasing function of $\theta_{p2}$, but when comparing categories 4 and 3  one has a decreasing function 
of $\theta_{p2}$. The behavior reflects that large values of $\theta_{p2}$ indicate the person's tendency to middle categories.

\section{Quantification of the Link between Person Parameters and the Ordering of Categories}

The definitions of ordinal models link  the latent trait to the probabilities of categories such that larger values of the trait tend to produce higher probabilities. Classical logistic models are ordinal according to all the definitions considered here. However, the differing concepts of ordinality show different aspects of the model and can be used to investigate the extent to which the ordering of categories is reflected by a specific  model. 

\subsection{Comparing Defining Functions}

Let us consider the cumulative model  as an example. Then all the functions
\[
g_r^{\text{split}}(\theta_p)=P(Y_{p} \ge r |  \theta_p) =F(\theta_p-\delta_r)
\]
are increasing in $\theta_p$. For the logistic model the function $g_r(\theta_p)$ is a logistic function  centered at $\delta_r$, that means, $g_r(\theta_p) =0.5$ if $\theta_p=\delta_r$. All the functions have the same form and increase from zero to one.

\begin{figure}[h!]
\begin{center}
\includegraphics[width=4.5cm]{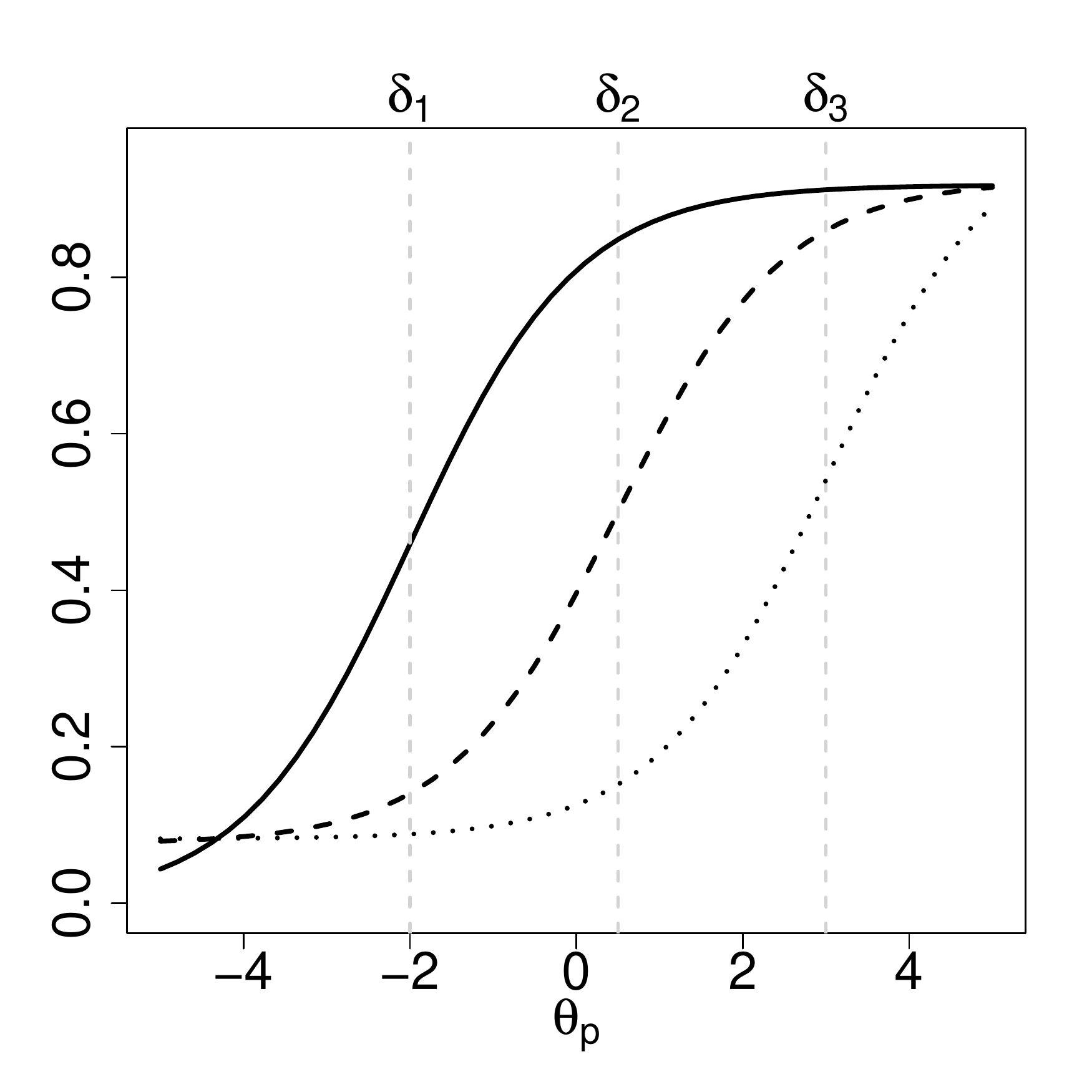}\includegraphics[width=4.5cm]{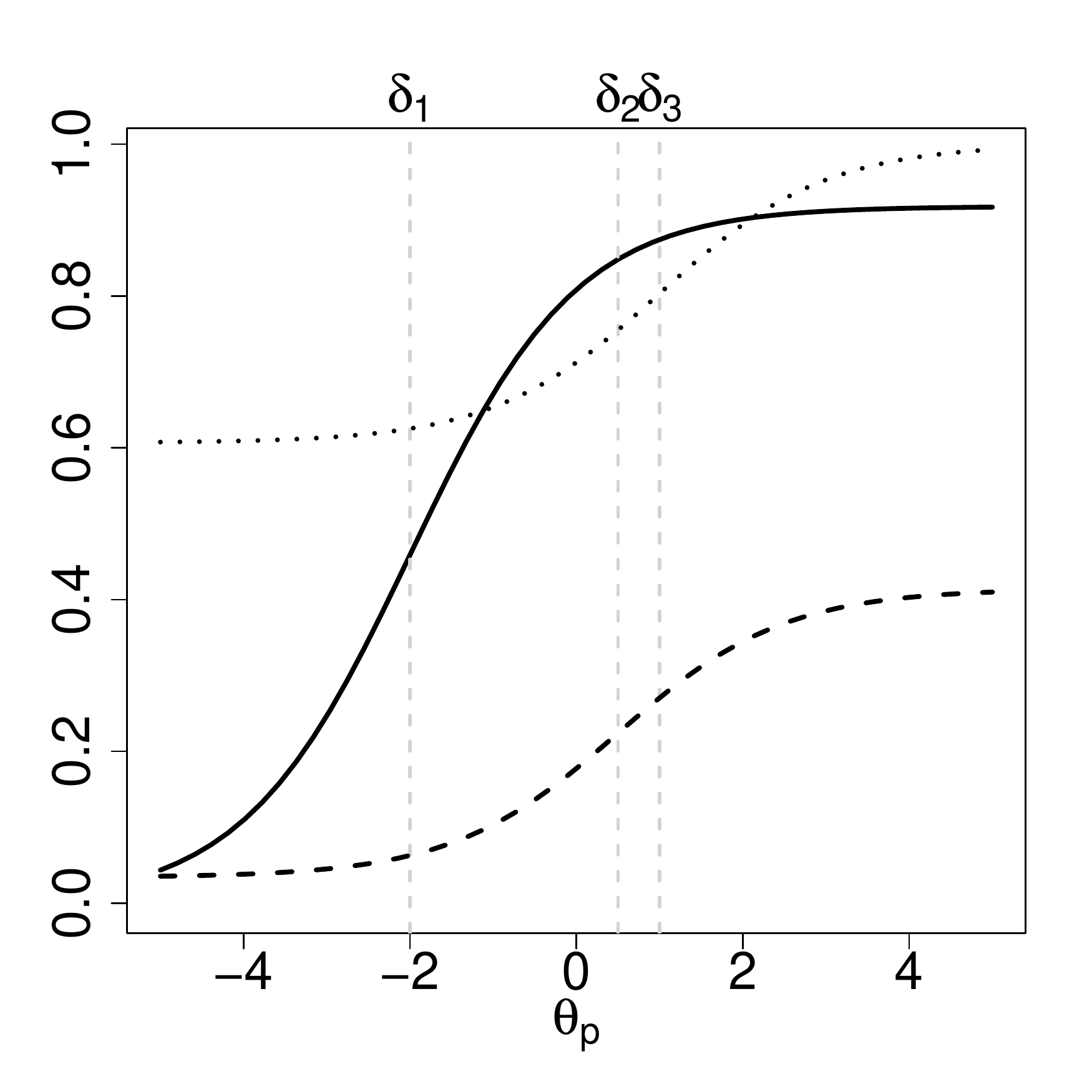}\includegraphics[width=4.5cm]{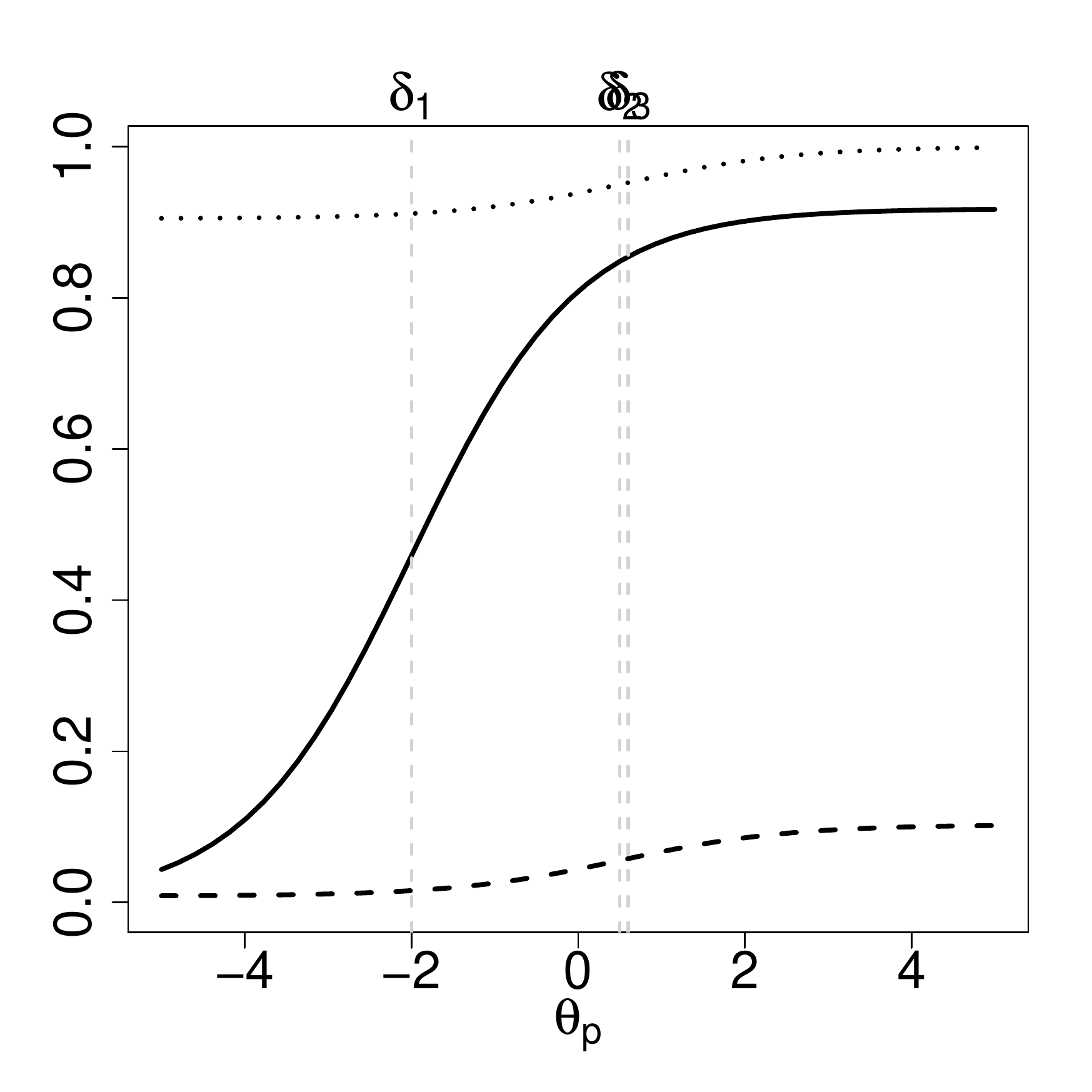}
\caption{Functions $g_r^{\text{adj}}(\theta_p)$, which compare categories 0,1 (drawn line), categories 1,2 (dashed line) and categories 2,3 (dotted line) for model with four categories }\label{fig:cutoff2}
\end{center}
\end{figure}

\begin{figure}[h!]
\begin{center}
\includegraphics[width=4.5cm]{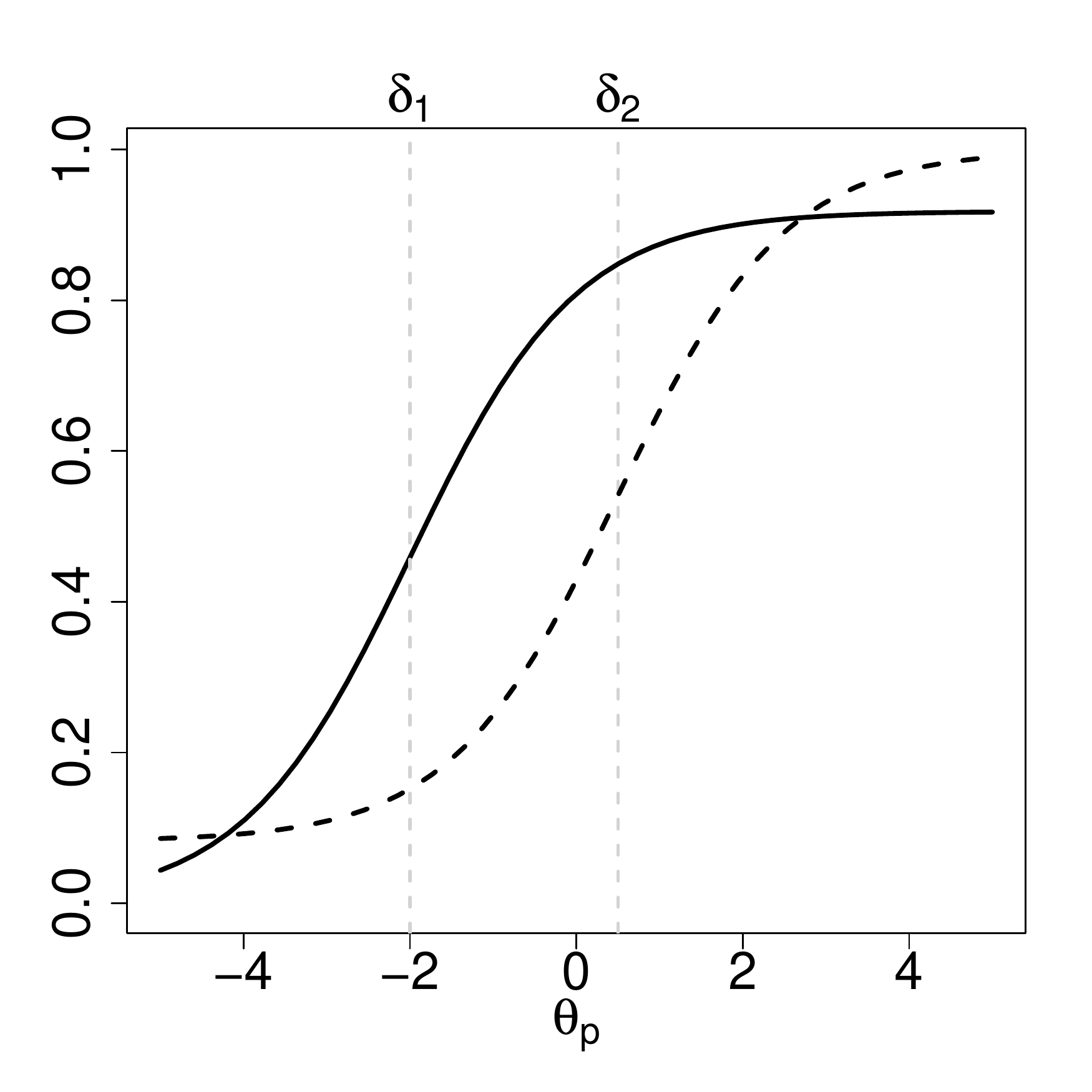}
\caption{Functions $g_r^{\text{adj}}(\theta_p)$,  which compare categories 0,1 (drawn line), categories 1,2 (dashed line) for model with three categories   after collapsing }\label{fig:cutoff3}
\end{center}
\end{figure}

For the model with logistic response function also the functions 
\[
g_r^{\text{adj}}(\theta_p)=P(Y_{p} = r | Y_{pi} \in \{r-1,r\}, \theta_p),
\]
which compare adjacent categories, are increasing in $\theta_p$. They are interesting since they show how well adjacent categories are separated by the model.
The strength of separation is determined by the item parameters. Figure \ref{fig:cutoff2} shows the functions $g_r^{\text{adj}}(\theta_p)$ for data generated from a cumulative model with logistic distribution function. The drawn line compares categories 0,1 , the dashed line  categories 1,2,  and the dotted line categories 2,3. It is seen from the left picture that for well separated thresholds the functions are similar, but shifted. If the thresholds $\delta_1, \delta_2$ 
are closer the functions that separate categories 1,2 and 2,3 are  flatter, in the right picture they have hardly any variation. Thus the trait has hardly an effect regarding the separation of the corresponding categories.   This may be seen as an indicator that one can collapse either by fusing categories 1 and 2 or by fusing categories 3 and 2. The effect is seen from Figure \ref{fig:cutoff3}, which shows that after collapsing the new categories 0,1 and 1,2 are well separated, the trait distinguishes well between the newly built categories. 

Collapsing of categories is in particular interesting in the  cumulative model since the model still holds after adjacent categories have collapsed. This is not true for other models as, for example, the partial credit model. If a partial credit model holds for data,   collapsing of adjacent categories destroys the model structure. All parameters change and the model does not necessarily hold after collapsing.

\subsection{Population Based Measures}

The strength of the link between person parameters and the ordering of categories can also be investigated with reference to the population for which items are supposed to work. 

Let $g_r(\theta_p)$ denote one of the ordinal functions that compares categories. Then a measure for the strength of the link between the person trait and the function is 
\begin{equation}
m_{g_r}= \int g_r'(\theta_p)f(\theta_p)d\theta_p,
\end{equation}
where $g_r'(\theta_p)$ denotes the derivative of $g_r(\theta_p)$ and $f(.)$ is the density of $\theta_p$ in the population. For the density one can typically assume a normal distribution. For a constant function $g_r(\theta_p)=c$ one obtains immediately $m_{g_r}=0$, indicating that there is no link between the latent trait and the function $g_r(\theta_p)$. Since for ordinal models $g_r'(\theta_p)$ is non-negative $m_{g_r}$ is always positive.

\begin{figure}[h!]
\begin{center}
\includegraphics[width=4.5cm]{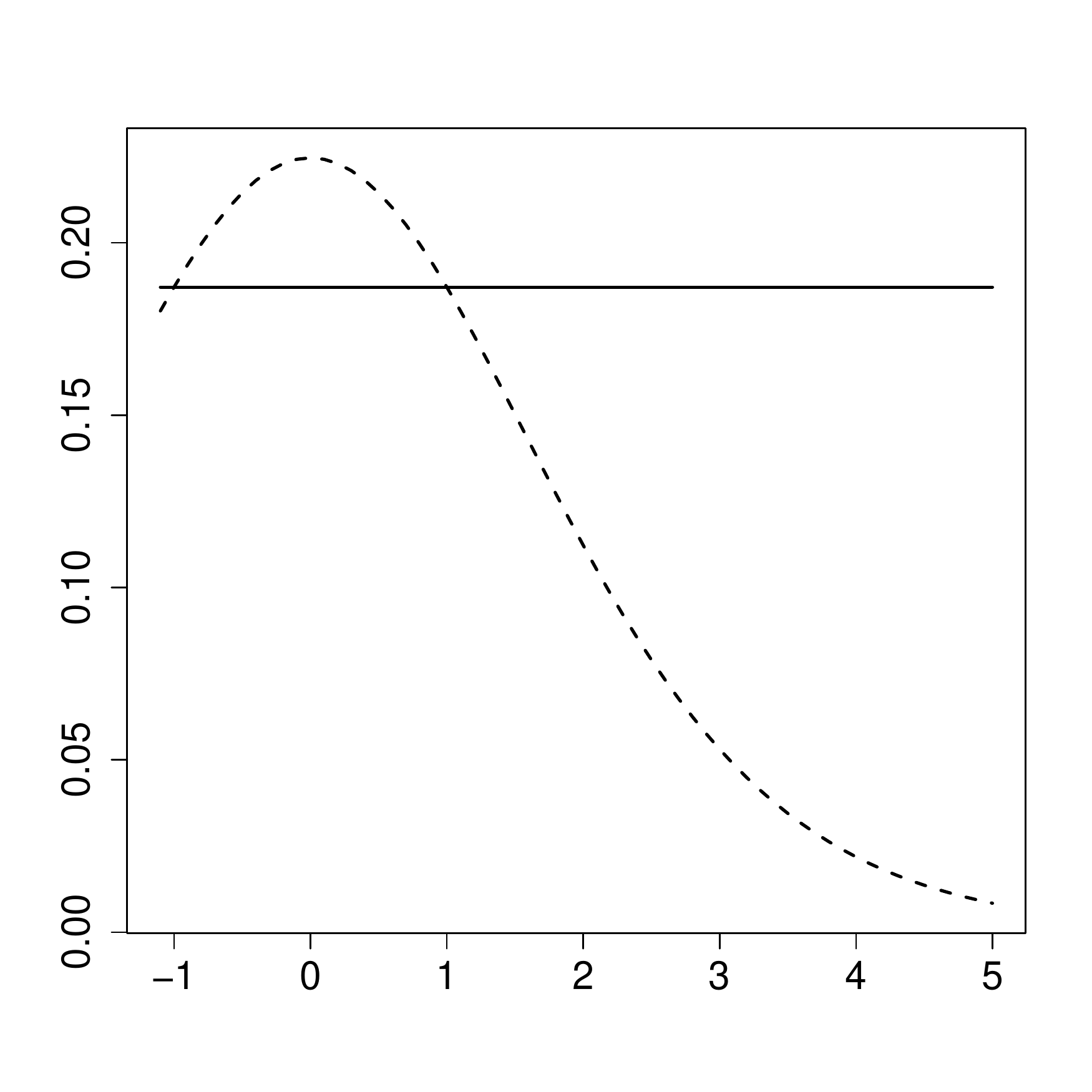}
\caption{Strength of  $g_r^{\text{adj}}(\theta_p)$ in a cumulative logit model as a function of the second threshold; first threshold is fixed at -1. Drawn line shows $m_1$, dashed line is $m_2$ }\label{fig:strength1}
\end{center}
\end{figure}

Let us consider again the split function $g_r^{\text{split}}(\theta_p)=P(Y_{p} \ge r |  \theta_p) =F(\theta_p-\delta_r)$. All the functions $g_r^{\text{split}}$, $r=1,\dots,k$ have the same shape, however show different strengths depending on the population. As an illustration let the first threshold in a cumulative logit model be fixed, $\delta_{i1}=-1$, and the second threshold $\delta_{i2}$ vary between -1 and 5. With only three response categories  and the population determined by a standard normal distribution, one obtains the  values of $m_{g_r}$ shown in Figure \ref{fig:strength1}.  
$m_{g_1}$ remains constant since the first threshold is fixed.  The strength $m_{g_2}$ is a function of the second threshold. It is seen that $m_2$ is large for small values of $\delta_{i2}$ but tends to zero when $\delta_{i2}$ increases. Thus, for large values the distinction between categories $\{0,1\}$ and $2$ seems pointless. In the population under consideration the link between the trait and the separation of these groups of categories is non existent. Consequently one might collapse categories 1 and 2 which are separated by the threshold $\delta_{i2}$. Of course, the main reason why $m_2$ is close to zero for large values of $\theta_p$ is that   category 2 is rarely observed in this population.

The strength described by $m_{g_r}$ refers to specific functions that compare categories. It should be noted that it provides no measure for the whole item.
A simple measure as the sum $m=m_{g_1}+\dots+m_{g_1}$ is certainly not   sensible   since it ignores that the separating functions ${g_r}$ are not independent. 

\section{Concluding Remarks}

Alternative definitions of ordinal models have been given, which coincide for classical logistic models. 
The increasing functions that are used do define ordinal models   reflect specific properties of items. They compare categories or groups of categories and, in the case of an ordinal model, provide stochastic ordering for persons. In particular they can be used to investigate  which comparison of categories is linked to the latent trait. In the multivariate case they also provide answers to the  question, which trait is linked to the ordering of categories.

The increasing functions reflect the ordering of categories, they  do not yield an ordering of items. This is a quite different problem that has been considered in the literature  on which we briefly comment. Ordering of items is   generally difficult in polytomous models since an item contains more than one parameter. There have been some attempts to define invariant item ordering, see \citet{sijtsma1998nonparametric}, \citet{hemker2001measurement}. They define that items have an \textit{invariant item ordering}   if they can be ordered and numbered such that 
\[
\E(Y_{p1}|\theta_p) \le \dots \le \E(Y_{pI}|\theta_p) \text{ for all } \theta_p,
\]
where $\E(Y_{pi}|\theta_p)$ is the expectation of the response of person $p$ on item $i$. However, as a measurement property it is questionable since computation of the  expectation assumes a higher scale level than ordinal. 
One obtains, for example,    that the  
sequential Rasch model does not imply invariant item ordering \citep{hemker2001measurement} but since the criterion is questionable the gain in knowledge is rather limited. 

What can be derived without assuming a higher scale level than is available  is invariant ordering of specific aspects of items. In the sequential Rasch model one can consider the functions
\[
g_{ r}^{\text{cond},i}(\theta_p)=P(Y_{pi} \ge r | Y_{pi} \ge r-1, \theta_p) =F(\theta_p-\delta_{ir}).
\]
If one focuses on the first step of the process of solving an item, which (for item $i$) is captured in $g_1^{\text{cond},i}(\theta_p)$, one obtains an invariant ordering of items with reference to the first step  in the form   
\[
g_{ 1}^{\text{cond},1}(\theta_p) \le \dots \le g_{ 1}^{\text{cond},I}(\theta_p) \text{ for all } \theta_p,
\]
after relabeling of items. In the same way one can compare other steps, however, one obtains no ordering of the   items as such.

\section*{Appendix}
\subsection{ Proposition \ref{th:gen}}
(1) Let the model be ordinal according to the paired concept, that is, for $\pi_{r}(\theta_p)=  P(Y_{p} = r|\theta_p)$ and all categories $s<r$
the functions $ \pi_{r}(\theta_p)/\pi_{s}(\theta_p)$  are increasing functions  of $\theta_p$.

Let us consider 
\begin{align*}
&P(Y_{p} \ge r | Y_{p} \ge r-1, \theta_p) =  \frac{P(Y_{p} \ge r|\theta_p)}{P(Y_{p} \ge r-1|\theta_p)}=\\
&= \frac{\pi_{r}(\theta_p)+\dots+\pi_{k}(\theta_p)}{\pi_{r-1}(\theta_p)+\pi_{r}(\theta_p)+\dots+\pi_{k}(\theta_p)}
= \frac{\pi_{r}(\theta_p)/\pi_{r-1}(\theta_p)+\dots+\pi_{k}(\theta_p)/\pi_{r-1}(\theta_p)}{1+\pi_{r}(\theta_p)/\pi_{r-1}(\theta_p)+\dots+\pi_{k}(\theta_p)/\pi_{r-1}(\theta_p)}=\\
&=F_{\text{log}} (\log(\pi_{r}(\theta_p)/\pi_{r-1}(\theta_p)+\dots+\pi_{k}(\theta_p)/\pi_{r-1}(\theta_p)))
\end{align*}
where $F_{\text{log}}(\eta)= \exp(\eta)/(1+\exp(\eta))$ is the logistic function.

Since all the functions $\pi_{s}(\theta_p)/\pi_{r-1}(\theta_p)$, $s\ge r$ are increasing functions  of $\theta_p$, and also $\log(.)$ and $F_{\text{log}}(.)$
are increasing functions  of $\theta_p$, the same holds for the function $P(Y_{p} \ge r | Y_{p} \ge r-1, \theta_p)$, which makes  the model   ordinal according to the conditional concept.                             

(2) Let the model be ordinal according to the conditional concept, that is,
$ P(Y_{p} \ge r | Y_{p} \ge r-1, \theta_p)$  is an increasing function  of $\theta_p$. 

For probabilities  the  general representation
\[
P(Y_{p} \ge r | \theta_p)= \prod^{r}_{s=1}P(Y_{p} \ge s | Y_{p} \ge s-1, \theta_p)
\]
holds. Thus, if all the functions on the right hand side, $ P(Y_{p} \ge r | Y_{p} \ge r-1, \theta_p)$, are increasing functions of $\theta_p$,
also the cumulative probability $P(Y_{p} \ge r | \theta_p)$ is an increasing function  of $\theta_p$, which makes the model ordinal according to the split concept.

\subsection{ Proposition \ref{df:cum2}}
Let the model be ordinal according to the split criterion. Thus, $F_r = P(Y_{p} \ge r|  \theta_p)$ is an increasing function of $\theta_p$.
To show that the model is ordinal according to the paired comparison  criterion one investigates for $s<r$
\[
\log (\pi_{r}(\theta_p)) - \log (\pi_{s}(\theta_p)).
\]
Tedious but straightforward derivation yields that for the logistic function $F(.)$ the derivative of this function is given by
\[
(\log (\pi_{r}(\theta_p)) - \log (\pi_{s}(\theta_p)))'= F(\theta_p-\delta_{r-1})-F(\theta_p-\delta_{r+1}),
\]
which is positive for any value of $\theta_p$. Therefore, the model is ordinal according to the paired comparison  criterion.

\bibliography{literatur}
\end{document}